\documentclass[aps,showkeys,showpacs,nofootinbib,floatfix]{revtex4}
\usepackage{amssymb}
\usepackage{amsfonts}
\usepackage{amsmath}
\usepackage{graphicx}
\usepackage{hyperref}
\usepackage{graphicx}

\begin{document}

\title{Self-gravitational Interaction in $z = 4$ Ho\u{r}ava-Lifshitz Gravity}
\author{Molin Liu$^{1}$}
\thanks{Corresponding author\\E-mail address: mlliu@xynu.edu.cn}
\author{Junwang Lu$^{1,2}$}
\author{Yin Xu$^{1}$}
\author{Jianbo Lu$^{2}$}
\author{Yabo Wu$^{2}$}
\author{Rumin Wang$^{1}$}
\affiliation{$^{1}$College of Physics and Electronic Engineering,
Xinyang Normal University, Xinyang, 464000, P. R. China\\
$^{2}$Department of Physics, Liaoning Normal University, Dalian, 116029, P. R. China}

\begin{abstract}
Motivated by recent works \cite{Rgcz4,Nobuyoshi}, the influences of self-gravitational interaction on the Hawking radiation are studied both for (3 + 1) and (4 + 1) dimensional black holes in $z = 4$ Ho\u{r}ava-Lifshitz gravity. It is found that the tunneling entropies $S_{B(3 + 1)}$ and $S_{B(4 + 1)}$ independent on particle's mass are consistent with the thermodynamical entropies $S_{BH(3 + 1)}$ and $S_{BH(4 + 1)}$, respectively. There is a very visible degree of uniformity between thermodynamics and quantum tunneling in $z = 4$ Ho\u{r}ava-Lifshitz gravity. It suggests that the entropies contained corrected terms could be explained well by the self-gravitational interaction of Hawking radiation. The study of tunneling process may shed light on understanding the Ho\u{r}ava-Lifshitz gravity.
\end{abstract}



\maketitle

\section{Introduction}
Recently, Ho\u{r}ava proposed a power-counting renormalizable Ho\u{r}ava-Lifshitz (HL) gravity at Lifshitz point \cite{Horavagravity}. It has completed properties in ultraviolet (UV) region with dynamical critical exponent $z$ under anisotropic Lifshitz scaling $t\rightarrow l^zt$, $\vec{x}\rightarrow l\vec{x}$. If the anisotropic scaling with $z \geq 3$ is realized in UV region, HL gravity is renormalizable. In order to keep the finite of quantum field, Lorentz invariance is given up in high energies, which leads foliation preserving diffeomorphism indicating invariant rather than full diffeomorphism. At large distance, it reduces to General Relativity (GR) with $z = 1$ fixed point. However, at short distance it represents a non-relativistic renormalizable gravity model with $z = 3$ fixed point. Since then $z = 3$ HL gravity attracts many attentions such as black hole solutions \cite{HLnewsolutionHJC,topological,HLnewsolutionKS,HLnewsolutionPark,HLnewsolutionother}, Hawking radiation \cite{LML1, LML2, peng}, thermodynamical properties \cite{HLrelixue}, perturbation \cite{LML2,QNM}, observational effects \cite{LMLGRG, observation, Konoplya, lensing}. About the overall reviews of HL gravity, one can refer to the papers \cite{reviewsHL}.

In early studies of Hawking radiation \cite{hawkingfaxian}, the background was fixed when particles radiate out from black holes \cite{HRclassmethod}. The pure thermal spectrum leads to the paradox of information loss and the broken down of the quantum unity \cite{yinan}. Soon after that, Wilczek and Kraus (WK) considered the self-interaction of gravity and treated radiation as a semiclassical tunneling process through Wentzel-Kramers-Brillouin (WKB) approximation \cite{kraus}. This method presented a possible interpretability to the paradox of information loss. Subsequently, Parikh and Wilczek \cite{ParikhWilczek} put forward the WK method with taking into account of both the energy conservation and the fixed Arnowitt-Deser-Misner (ADM) mass of the system, named as geodesic method or Parikh-Wilczek (PW) method. In PW method, the potential barrier needed in quantum tunneling picture is obtained by radiating particles, and the mass of black holes decrease for Hawking radiation.Whereafter, Hawking radiation of various black holes are calculated by PW method \cite{tunnelingworks}. These models of Hawking radiation as tunneling show the actual spectrum, related to the variety of the Beikenstein-Hawking entropy, deviates from the pure thermodynamics in the tunneling process.

After Ho\u{r}ava gave out original HL gravity, Cai, Liu and Sun proposed the $z = 4$ HL gravity in Ref.\cite{Rgcz4} which realizes power counting super renormalizbale in UV region successfully. The calculation of spectrum dimension becomes more meaningful than the causal dynamical triangulation (CDT) approach. However, it is very interesting that some modifications   appear in the original $z = 4$ HL gravity \cite{Rgcz4}. In Cai and Ohta's casting model, these new entropy as well as the mass of $z = 4$ HL gravity are obtained naturally by matching gravitational field equation and the first law of thermodynamics like in Einstein gravity or Lovelock gravity \cite{Nobuyoshi}. As far as we known the explanations about these new entropies are only above cast model in Ref.\cite{Nobuyoshi}. Fortunately, the logarithmic entropy of Kehagias-Sfetsos black hole is explained well by the tunneling with self-gravitational interaction of asymptotically flat IR modified HL gravity \cite{LML1}. Hence, motivated by above various factors, Hawking radiation as  tunneling is studied in $z = 4$ HL gravity here.

This paper is organized as followings. In Section II, we briefly present $z = 4$ HL gravity. In Section III, we use geodesic method to study the massless and massive particle's tunneling for the (3 + 1) dimensional black hole in $z = 4$ HL gravity. In Section IV, we study the tunneling for the (4 + 1) dimensional black hole in $z = 4$ HL gravity. Section V is the conclusion. We adopt the signature ($-$, $+$, $+$, $+$) and put $\hbar$, $c$, and $G$ all equal to unity.
\section{(3 + 1) and (4 + 1) dimensional black holes in $z = 4$ Ho\u{r}ava-Lifshitz Gravity}
In this section, we briefly review $z = 4$ HL gravity \cite{Rgcz4}. By using the ADM decomposition, the metric of the (D+1) dimensional spacetime is written as,
\begin{equation}
ds^2 = -N^2 dt^2 + g_{ij}\left(dx^i - N^{i}dt\right)\left(dx^j - N^{j}dt\right),
\end{equation}
with $i = 1,\ 2,\ 3,\ \ldots, D$. Because the dimensions of time and space in the HL gravity are $[t] = -z$, $[x] = -1$, $[c] = z - 1$ at the fixed point with Lifshitz index $z$, this HL gravity reduces to GR in the IR limit when $z = 1$. In UV limit, this type HL gravity chooses other index $z$ to make itself renormalizable. When $z_{UV} = D$, the theory is power-counting renormalizable. While $z_{UV} > D$, this theory becomes super-renormalizable. Lapse function $N$, shift function $N_i$, spatial metric $g_{ij}$ are the functions of $t$ and $x$. Considering the simplest case, the lapse function $N$ is restricted to depend only on the time coordinate $t$. In fact, this requirement imposed on $N$ is the projectability condition which results from the foliation preserving diffeomorphism.

The kinetic term with constant $\kappa$ and $\lambda$ can be constructed as followings,
\begin{equation}
S_K=\frac{2}{\kappa^2}\int dtd^Dx\sqrt{g}N\left(K_{ij}K^{ij}-\lambda K^2\right),\label{1kinetic}
\end{equation}
where $K_{ij}=\frac{1}{2N}\left(\dot{g_{ij}}-\nabla_{i}N_j-\nabla_{j}N_i\right)$  is the second fundamental form of extrinsic curvature, and an over dot denotes the derivative with respect to $t$. According to the dimension of volume element $[dt d^D x] = - D - z$ and the time derivative dimension $[\partial_t]=z$, the coupling constant $\kappa$\ has dimension $[\kappa] = (z-D)/2$. Here, $\lambda$ is dimensionless parameter which is susceptible to the quantum correction \cite{Horavagravity}. If $\lambda=1$ is taken in the IR region, the theory can reduce to GR.

According to the detail balance condition, the potential term can be formed as followings,
\begin{equation}
S_V=\frac{\kappa^2}{8}\int dtd^Dx\sqrt{g}NE^{ij}\zeta_{ijkl}E^{kl},\label{1potential}
\end{equation}
where $E^{ij}=\frac{1}{\sqrt{g}}\frac{\delta W_D[g_{ij}]}{\delta g_{ij}}$ is detail balance condition connecting the action $W_D$ of D dimensional system and the potential terms of $D+1$ dimensional system. $\zeta _{ijkl}=\frac{1}{2}\left(g_{ik}g_{jl}+g_{il}g_{jk}\right)-\tilde{\lambda}g_{ik}g_{jl}$ with constant $\tilde{\lambda}=\frac{\lambda}{D\lambda-1}$ is the inverse of the generalized De Witt metric which defines the form quadratic in $K_{ij}$. Considering the kinetic term (\ref{1kinetic}) and potential term (\ref{1potential}), the (3 + 1) and (4 + 1) dimensional actions of the $z = 4$ HL gravity are obtained in Ref.\cite{Rgcz4}. For the need of concise statement, we do not list them in this paper, about the details please refer to original Ref.\cite{Rgcz4}.

Based on $z = 4$ HL gravity actions \cite{Rgcz4}, the black holes solutions are obtained, thereby, and shown as
\begin{equation}\label{duguijiashe7}
ds_{D + 1}^2=- f_{D + 1}(r) dt^2+\frac{dr^2}{f_{D + 1}(r)} + r^2 d\Omega_{k(D + 1)}^2.
\end{equation}
For the (3 + 1) dimensional case, $d\Omega_{k(3 + 1)}^2$ represents the line element of $2$-dimensional Einstein space with a constant curvature $2k$. Here, in the interest of simplicity $\lambda=1$ and $\beta=-3/8$ are taken in the IR region. The lapse function $f_{3 + 1}(r)$ is written as
\begin{equation}\label{solution4}
f_{3 + 1}=k+\frac{\tilde{\mu}}{2\tilde{\beta}}x^2
\left(1-\sqrt{1-\frac{4\tilde{\beta}}{x^2\tilde{\mu}^2}\left(\tilde{\mu}x^2-\sqrt{c_0x}\right)}\right),
\end{equation}
where $\tilde{\mu}=-\mu\Lambda_W$, $\tilde{\beta}=\frac{\Lambda_W^2}{4M}$, $x=\sqrt{-\Lambda_W}r$ and $c_0$ is a positive integration constant. The solution (\ref{solution4}) is the ``$-$" branch of the actual full solutions because when $\tilde{\beta}\rightarrow 0$ it should return to the $z = 3$ HL gravity \cite{Rgcz4}. Then, by using the Hamiltonian method and the first law of thermodynamics \cite{Hamilton, topological}, the (3 + 1) dimensional black hole's mass $M_{3 + 1}$, temperature $T_{3 + 1}$,  thermodynamical entropy $S_{BH(3 + 1)}$ are given by,
\begin{eqnarray}
  M_{3 + 1} &=& \frac{\kappa^2\Omega_k}{16(-\Lambda_W)^{\frac{3}{2}}}c_0,\ \
c^{\prime}=\frac{16(-\Lambda_W)^{\frac{3}{2}}}{\kappa^2\Omega_k},\ \ c_0=c^{\prime}M_{3 + 1}, \label{shi9}\\
T_{3 + 1} &=& \frac{\sqrt{-\Lambda_W}}{8\pi x_+}
\frac{\left(3x_+^2 - k\right)\tilde{\mu}x_+^2 - 5\tilde{\beta}k^2}{\tilde{\mu}x_+^2 + 2\tilde{\beta}k}, \label{adt31} \\
S_{BH(3 + 1)} &=& \frac{1}{4}\pi\kappa^2\mu^2\Omega_{k(3 + 1)}\left(x_+^2+2k\ln x_+-\frac{3\tilde{\beta}k^2}{\tilde{\mu}x_+^2}-
\frac{\tilde{\beta}^2k^3}{\tilde{\mu}^2x_+^4}+\frac{4\tilde{\beta}k}{\tilde{\mu}}\ln x_+\right)+S_0,\label{shi11}
\end{eqnarray}
where $S_0$ is an integration constant and $r_+$ is event horizon. It is interesting that except for the quadratic term there are many corrections in entropy (\ref{shi11}) comparing the  Beikenstein-Hawking area entropy in usual GR.

For the (4 + 1) dimensional case, $d\Omega_{k(4 + 1)}^2$ denotes the element line of $3$-dimensional Einstein's space manifold which possesses an arbitrary constant curvature $6k$. In order to simplify the computation, $\beta=-1/3$, $\lambda=1$ are adopted here. According to the action, the lapse function $f_{4 + 1}$ is given by
\begin{equation}\label{equation14}
f_{4 + 1}=k+\frac{x^2}{3}-\sqrt{c_0},
\end{equation}
where $c_0$ is an integration constant and $x = \sqrt{-\Lambda_W}r$. This type solution comes from the ``$-$" branch of total solutions by comparing with the Gauss-Bonnet gravity \cite{Rgcz4}. The corresponding (4 + 1) dimensional black hole's mass $M_{(4 + 1)}$, temperature $T_{(4 + 1)}$, thermodynamical entropy $S_{BH(4 + 1)}$ are listed as \cite{Rgcz4},
\begin{eqnarray}
M_{4 + 1} &=& \frac{\kappa^2\mu^2\Omega_{k(4 + 1)}}{24}c_0=\frac{c_0}{c^{\prime\prime}},\ \ c^{\prime\prime} = \frac{24}{\kappa^2\mu^2 \Omega_{k(4 + 1)}},\label{equation15}\\
T_{4 + 1} &=& \frac{\sqrt{-\Lambda_W}}{6\pi}x_+,\label{addt41} \\
S_{BH(4 + 1)} &=& \frac{\pi\kappa^2\mu^2\Omega_{k(4 + 1)}}{27\sqrt{-\Lambda_W}}(x_+^3 + 9kx_+) + S_0. \label{equation17}
\end{eqnarray}
It is also interesting found that there is not any quadratic term in entropy (\ref{equation17}), which departs from the pure Beikenstein-Hawking area entropy significantly.
\section{self-gravitational interaction for (3 + 1) dimensional black holes in $z = 4$ Ho\u{r}ava-Lifshitz gravity}
In this section, we study the influence of self-gravitational interaction on the Hawking radiation for (3 + 1) dimensional black holes in $z = 4$ HL gravity by using geodesic method. First of all, by using the Painlev\'{e} transformation \cite{ParikhWilczek,Painleve}, the original metric (\ref{duguijiashe7}) is rewritten as,
\begin{equation}
ds_{D + 1}^2=-f_{D + 1}(r, M)dt^2+2\sqrt{1-f_{D + 1}(r, M)} dt dr + dr^2 + r^2 d\Omega_{k(D + 1)}^2,\label{duguijiashe19}
\end{equation}
where $t$ is the Painlev\'{e} type time, $D = 3$ and $4$ are corresponding to (4 + 1) dimensions and (3 + 1) dimensions, respectively. Although Painlev\'{e} transformation is used, the fundamental symmetry of HL gravity is not violated because the metric (\ref{duguijiashe19}) could be looked like obtained via the followings foliation preserving diffeomorphisms maps: $d t\longrightarrow d t$, $dr \longrightarrow \sqrt{2 f_{D + 1} \sqrt{1 - f_{D + 1}}d r dt + f_{D + 1} d r^2}$.

According to the WKB approximation, one can obtain the relation between the imaginary part of action $S_{p(D + 1)}$ and the emission rate $\Gamma$ shown by,
\begin{equation}\label{shi20}
\Gamma\sim e^{-2Im S_{p(D + 1)}}.
\end{equation}
In this paper, the calculation of $Im S_{p(D + 1)}$ is divided into two parts: one is related to the massless particles case and another is related to the massive particles case. We first calculate the massless case. According to the null geodesic $ds^2 = 0 = d\Omega_{k(D + 1)}$ with Painlev\'{e} type metric (\ref{duguijiashe19}), the radial geodesic equation is given by
\begin{equation}\label{equation21}
\dot{r}=\pm1-\sqrt{1 - f_{D + 1}(r, M)},
\end{equation}
where $+$ ($-$) indicates the outgoing (incoming) particles. Here, only the outgoing case is considered. The particles created outside the event horizon with negative energy is tunneling  into the event horizon. However, the time need to be reverse transformed, i.e. $g_{01}\rightarrow -g_{01}$, because the negative energy particle propagates backwards in time. Meanwhile, negative energy particles can see that the black hole geometry is unchanged, the mass of black hole $M - \omega$ for the outgoing case should be replaced by the mass $M + \omega$ for ingoing particles taking into account of the self-gravitational interaction.

The imaginary part of radial outgoing particles' action is written as
\begin{equation}\label{S22}
Im S_{p(D + 1)} = Im \int_{r_i}^{r_f}P_rdr = Im\int_{r_i}^{r_f} \int_{0}^{P_r}d{{p}^\prime_r}dr,
\end{equation}
where $r_i$ and $r_f$ are the locations of horizons before and after tunneling happened, $p^\prime_r$ is the radial momentum. By using Hamiltonian equation $\frac{d H}{d p^\prime_r} = \frac{dr}{dt} = \dot{r}(M)$, $Im S_{p(D + 1)}$ (\ref{S22}) is rewritten in a solvable form as
\begin{equation}\label{equation24}
Im S_{p(D + 1)}=Im\int_{r_i}^{r_f} \int_{M}^{M-\omega}\frac{d(M-\omega)}{\dot{r}(M-\omega)}dr
=Im\int_{r_i}^{r_f} \int_{M}^{M-\omega}\frac{d(M-\omega)}{1-\sqrt{1-f_{D + 1}(r,M-\omega)}}dr,
\end{equation}
where the self-gravitational interaction is considered as followings. The mass of black hole has to change as $M \rightarrow (M - \omega)$ after the particle radiating out from the black hole. Naturally, lapse function $f_{D + 1}(r,M)$ and geodesic equation $\dot{r}(M)$ change to $f_{D + 1}(r,M-\omega)$ and $\dot{r}(M-\omega)$, respectively. In order to simplify calculating we adopt a new variable as
\begin{equation}\label{addalpha}
\alpha = \sqrt{1-f_{D + 1}}.
\end{equation}

Then, the mass of (3 + 1) dimensional black hole is shown by
\begin{equation}\label{331m1c}
M = \frac{1}{c^{\prime}}\left[\tilde{\beta}x^{-\frac{5}{2}}(\alpha^2 - 1 + k)^2 + \tilde{\mu}x^{-\frac{1}{2}}(\alpha^2 - 1 + k) +
\tilde{\mu}x^{\frac{3}{2}}\right]^2,
\end{equation}
where $\alpha = \sqrt{1-f_{3 + 1}}$ is for (3 + 1) dimensions and $c^{\prime}$ is shown by Eq.(\ref{shi9}). Then according to the energy conservation which means that the black hole's mass decreases from $M$ to $M - \omega$ after a particle with energy $\omega$ tunneling out, the derivative of $M$ with respect to $\alpha$ is shown by
\begin{equation}\label{shi25}
d(M-\omega) = \frac{2}{c^{\prime}}\left(P\alpha^7+Q\alpha^5+m\alpha^3+n\alpha\right)d\alpha,
\end{equation}
where $P$, $Q$, $m$, $n$ are listed as followings,
\begin{eqnarray}
\nonumber P&=&4\tilde{\beta}^2x^{-5},\label{equation2610}\\
\nonumber Q&=&12\tilde{\beta}^2x^{-5}(k-1)+6\tilde{\beta}\tilde{\mu}x^{-3},\label{equation261}\\
\nonumber m&=&12\tilde{\beta}^2x^{-5}(k-1)^2+6\tilde{\beta}\tilde{\mu}x^{-3}(k-1)+4\tilde{\beta}\tilde{\mu}x^{-1}
+2\tilde{\mu}^2x^{-1},\label{equation262}\\
\nonumber n&=&4\tilde{\beta}^2x^{-5}(k-1)^3+6\tilde{\beta}\tilde{\mu}x^{-3}(k-1)^2+(4\tilde{\beta}\tilde{\mu}x^{-1}
+2\tilde{\mu}^2x^{-1})(k-1)+2\tilde{\mu}^2x.\label{equation26}
\end{eqnarray}
Hence, after considering the self-gravitational interaction, $Im S_{p(3 + 1)}$ (\ref{equation24}) is rewritten as
\begin{eqnarray}
 Im S_{p(3 + 1)} &=& -\frac{2}{c^{\prime}}Im \int_{r_i}^{r_f} \int_{\alpha_i}^{\alpha_f}\left[
\mathcal{F}_{01}(\alpha) + \mathcal{F}_1(\alpha) \right]d\alpha dr,\label{equation27}\\
\mathcal{F}_{01}(\alpha) &=& P(\alpha^6 + \alpha^5) + (P + Q)(\alpha^4 + \alpha^3) + (P + Q + m)(\alpha^2 + \alpha) + P + Q + m + n,\label{311f0}\\
\mathcal{F}_1(\alpha) &=& \frac{P + Q + m + n}{\alpha - 1},\label{311f01}
\end{eqnarray}
where $\alpha \in [\alpha_i,\alpha_f]$ is corresponding to the mass in the range of $[M, M - \omega]$. Obviously, in the integration the function of $\mathcal{F}_{01}(\alpha)$ (\ref{311f0}) is analytic in the range $[\alpha_i,\alpha_f]$. However, the last term $\mathcal{F}_1(\alpha)$ (\ref{311f01}) has a singularity at the point $\alpha = 1$ in the same domain, which is
corresponding to the event horizon position of black hole in $z = 4$ HL gravity. Hence, $\mathcal{F}_1(\alpha)$ becomes the only one contributed to the imaginary part of the action. Due to the singularity in the definition domain, we can only integrate Eq.(\ref{equation27}) in the complex plane. Then adopting the contour integration on the upper half of complex plane $\alpha$, we can get
\begin{equation}
Im S_{p(3 + 1)} = -\frac{2\pi}{c^{\prime}}\int_{r_i}^{r_f}\left(4k^3\tilde{\beta}^2x^{-5} +
6k^2\tilde{\beta}\tilde{\mu}x^{-3} +
4k\tilde{\beta}x^{-1} + 2k\tilde{\mu}^2x^{-1} + 2\tilde{\mu}^2x\right)dr,\label{shi28}
\end{equation}
where $x = \sqrt{-\Lambda_W}r$ and a replacement $\alpha - 1 = \rho e^{i\xi}$ is used with $\xi\in [0, \pi]$.

After obtaining the massless particles' $Im S_{p(3 + 1)}$ (\ref{shi28}), we turn to the aspect of massive case. As mentioned before in Jiang-Wu-Cai's paper \cite{Qingquanjiang}, the motion of massive particle is influenced by the Lorentz force which leads directly the particle does not move along the null geodesic. The trajectory of radiating particles are no longer light-like. However, if we treat the radiating particles near horizon as the spherical De Broglie wave (S wave), the geodesic of radiating particles can be looked as the group velocity of the de Broglie wave \cite{zhang111}.

For a de Broglie wave, according to the definition of group velocity $v_g$ and phase velocity $v_p$, we can get $v_p = v_g/2$. Based on the clock synchronization in Landau's theory \cite{Landau112}, the geodesic equation of massive particles is given by,
\begin{equation}\label{333dotr}
\dot{r} = v_p = \frac{1}{2}v_g = -\frac{1}{2}\frac{g_{tt}}{g_{tr}},
\end{equation}
where we adopt the Painlev\'{e} type metric (\ref{duguijiashe19}). If we consider the energy conservation, the imaginary part of the massive particles action is written as
\begin{equation}\label{equation35}
Im S_{p(D + 1)} = Im\int_{r_i}^{r_f} \int_{M}^{M-\omega}\frac{2\sqrt{1-f_{D + 1}(r,M-\omega)}
d(M-\omega)}{f_{D + 1}(r,M-\omega)}dr.
\end{equation}

For the case of (3 + 1) dimensional black hole, after considering the self-gravitational interaction, the imaginary part of massive particle is given by
\begin{eqnarray}
Im S_{p(3 + 1)} &=& -\frac{2}{c^{\prime}}Im \int_{r_i}^{r_f} \int_{\alpha_i}^{\alpha_f}\left[\mathcal{F}_{02}(\alpha) - \mathcal{F}_{2}(\alpha) + \mathcal{F}_{1}(\alpha)\right]d\alpha dr,\label{equation37}\\
\mathcal{F}_{02}(\alpha)&=& 2\left[P \alpha^6 + (P + Q)\alpha^4 + (P + Q + m)\alpha^2 + P + Q + m + n\right], \label{332mathf02}\\
\mathcal{F}_{2}(\alpha)&=& \frac{P + Q + m + n}{\alpha+1},\label{332mathf22}
\end{eqnarray}
where $\mathcal{F}_{1}(\alpha)$ is already presented in Eq.(\ref{311f01}). Obviously, $\mathcal{F}_{02}(\alpha)$ (\ref{332mathf02}), is analytic in the domain $[\alpha_i,\alpha_f]$. Although, there is a singularity point at $\alpha=-1$ in $\mathcal{F}_{2}(\alpha)$ (\ref{332mathf22}), but this point is out of the limit of integration. Hence, $\mathcal{F}_{02}(\alpha)$ and $\mathcal{F}_{2}(\alpha)$ have no contribution to the imaginary part of the particles' action. So only the last term $\mathcal{F}_{1}(\alpha)$ (\ref{311f01}) determines the imaginary part of action. Considering above situation, the results of massive particles are the same as the former massless case. The tunneling rate is the same form, namely Eq.(\ref{shi28}), for all type tunneling no matter with mass or not.

Finally, according to the emission rate (\ref{shi20}) and the imaginary part (\ref{shi28}), which are both applicable to massless and massive particles, we can obtain the total tunneling rate for the (3 + 1) dimensional black hole in $z = 4$ HL gravity written as
\begin{equation}\label{equation31}
\Gamma\sim e^{-2Im S_{p(3 + 1)}} = e^{(S_{bf}-S_{bi})} = e^{\Delta S_{B(3 + 1)}},
\end{equation}
where $S_{bf}$ denotes the entropy at the event horizon of black hole from whom particles have radiated out, $S_{bi}$ represents the entropy at the horizon of black hole before particles radiate out. $\Delta S_{B(3 + 1)}$ is the change of tunneling entropy which leads directly a non-area entropy shown by
\begin{equation}\label{equation32}
S_{B(3 + 1)} = \frac{1}{4}\pi\kappa^2\mu^2\Omega_{k(3 + 1)}\left(x^2 + 2k\ln x - \frac{3k^2\tilde{\beta}}{\tilde{\mu}x^2} - \frac{k^3\tilde{\beta}^2}{\tilde{\mu}^2x^4} +
\frac{4k\tilde{\beta}}{\tilde{\mu}}\ln x\right).
\end{equation}
In this quantum tunneling of (3 + 1) dimensional black hole, the tunneling entropy $S_{B(3 + 1)}$ proves that the spectrum of Hawking radiation has relation to the change of entropy at the horizon before and after Hawking radiation. It is also interesting that this tunneling entropy obtained through PW method is the same as the thermodynamical entropy (\ref{shi11}) obtained via the first law of thermodynamics. The entropy of HL gravity maybe is not in a horizon area form simply. If we consider the self-gravitational interaction of radiating particles in $z = 4$ HL gravity, the actual spectrum is not the pure thermodynamics but rather the spectrum related to the change of entropy of the system.
\section{self-gravitational interaction for (4 + 1) dimensional black holes in $z = 4$ Ho\u{r}ava-Lifshitz gravity}
In this section, we continue to study the influence of self-gravitational interaction on the Hawking radiation through the geodesic tunneling method but for (4 + 1) dimensional black holes in $z = 4$ HL gravity.

Firstly, the mass of (4 + 1) dimensional black hole is shown by
\begin{equation}\label{add41dimension}
M = \frac{1}{c^{\prime\prime}}\left(\alpha^2 - 1 + k - \frac{1}{3}\Lambda_W r^2\right)^2,
\end{equation}
where $\alpha = \sqrt{1 - f_{4 + 1}}$. For the tunneling of massless particle, the radial light-like geodesic equation in (4 + 1) dimensional black hole is given by Eq.(\ref{equation24}) with self-gravitational interaction. According to the (4 + 1) dimensional Painlev\'{e} type metric (\ref{duguijiashe19}) and the energy conservation after considering self-gravitational interaction, the imaginary part of radiating massless particles' action is obtained thereby,
\begin{eqnarray}
Im S_{p(4 + 1)} &=&-Im\frac{4}{c^{\prime\prime}}\int_{r_i}^{r_f} \int_{\alpha_i}^{\alpha_f}\left(\mathcal{F}_{03} + \mathcal{F}_3\right)d\alpha dr,\label{equation45}\\
\mathcal{F}_{03} &=& \alpha^2 + \alpha + k - \frac{1}{3}\Lambda_W r^2,\label{421mathf03}\\
\mathcal{F}_3 &=& \frac{k - \frac{1}{3}\Lambda_W r^2}{\alpha-1},\label{421mathf3}
\end{eqnarray}
where $c^{\prime\prime}$ is shown by Eq.(\ref{equation15}).

Obviously, $\mathcal{F}_{03}$ (\ref{421mathf03}) is analytic in the domain $[\alpha_i,\alpha_f]$. Hence, only $\mathcal{F}_3(\alpha)$ (\ref{421mathf3}) effects the imaginary part of the action (\ref{equation45}). Through the complex plane integral of above formula (\ref{equation45}), the
final imaginary part of massless particle is obtained as
\begin{equation}
Im S_{p(4 + 1)} = -\frac{1}{6}\pi\kappa^2\mu^2\Omega_{k(4 + 1)}\left[\left(k r_f-\frac{1}{9}\Lambda_W r_{f}^3\right)-\left(k r_i-\frac{1}{9}\Lambda_W r_{i}^3\right)\right].
\end{equation}

Secondly, the imaginary part for (4 + 1) dimensional massive particles could be also given out like before. With the help of the group velocity, the phase velocity and the landau's condition of the coordinate clock synchronization \cite{zhang111,Landau112}, one can obtain the geodesic equation of radial outgoing massive particles thereby. Then according to Eq.(\ref{equation35}) and considering the self-gravitational interaction, the imaginary part of massive particles' action is given by
\begin{eqnarray}\label{equation52}
Im S_{p(4 + 1)}&=&-\frac{4}{c^{\prime}}Im\int_{r_i}^{r_f} \int_{\alpha_i}^{\alpha_f}\left[\mathcal{F}_{04} (\alpha) - \mathcal{F}_4 (\alpha) + \mathcal{F}_3 (\alpha) \right]d\alpha dr,\\
\mathcal{F}_{04}(\alpha) &=& \alpha^2 + k - \frac{2}{3}\Lambda_W r^2,\label{42mathf04}\\
\mathcal{F}_4 (\alpha) &=& \frac{k - \frac{1}{3}\Lambda_W r^2}{\alpha + 1},\label{42mathf4}
\end{eqnarray}
where $\mathcal{F}_{04}(\alpha)$ is analytic in the domain $[\alpha_i, \alpha_f]$. Meanwhile, although there is a singularity point at $\alpha=-1$ in the second term $\mathcal{F}_{4}(\alpha)$, this point is out of the integrating range. Hence, $\mathcal{F}_{04}(\alpha)$ and $\mathcal{F}_{4}(\alpha)$ have no contribution to the imaginary part of the particles' action. So only the last term $\mathcal{F}_3 (\alpha)$ (\ref{421mathf3}) works to Eq. (\ref{equation52}). Until now, it is very interesting that both for massless and massive particles they all have the same imaginary part of the tunneling particles' action, which will be used to calculate the tunneling rate and tunneling entropy.

According to the relationship between the tunneling rate and the imaginary part of particles' action (\ref{shi20}), the final tunneling rate of massless particles for (4 + 1) dimensional black hole in $z = 4$ HL gravity is given by
\begin{eqnarray}\label{equation47}
\Gamma = \exp \left[\frac{1}{27}\pi\kappa^2\mu^2\Omega_{k(4 + 1)}\left[\left(9k r_f-\Lambda_W r_{f}^3)-(9k r_i-\Lambda_W r_{i}^3\right)\right]\right] = e^{\Delta S_{B(4 + 1)}},
\end{eqnarray}
where $\Delta S_{B(4 + 1)}$ is the change of tunneling entropies. According to the change of tunneling entropy above we can get the final tunneling entropy of massless particles tunneling around (4 + 1) dimensional black hole in $z = 4$ HL gravity shown by
\begin{equation}\label{equation48}
S_{B(4 + 1)} = \frac{\pi\kappa^2\mu^2\Omega_{k(4 + 1)}}{27\sqrt{-\Lambda_W}}(9kx+x^3) + S_0,
\end{equation}
where $S_0$ is constant. It is interesting that this result $S_{B(4 + 1)}$ Eq.(\ref{equation48}) obtained via PW method is the same as the result (\ref{equation17}) obtained via the first law of thermodynamics at event horizon. It is known that $r_i$ satisfies $f_{4 + 1}(r_i, M)=0$. when a particle radiates out from the event horizon, the mass of black hole will decrease. So the new event horizon $r_f$ will satisfies $f_{4 + 1}(r_f, M-\omega)=0$. Obviously, the tunneling rate is related to the gap between $r_i$ and $r_f$. So the spectrum of Hawking radiation is not the pure thermodynamics but rather the pure thermodynamics spectrum with some deviation. Through above analysis, the non-area quantum tunneling entropy Eq.(\ref{equation48}) deviated from pure Beikenstein-Hawking area entropy are explained well by the self-gravitation in the tunneling picture. The radiation spectrums of $z = 4$ black hole are relevant to the entropy change before and after tunneling.
\section{Conclusion}
In this paper, we have studied the self-gravitational interaction in Hawking radiation as tunneling both for (3 + 1) and (4 + 1) dimensional black holes in $z = 4$ HL gravity. In this section we summarize what we have obtained.

1. There is one interesting phenomenon in most of HL gravities, namely the entropy of black hole deviates from the pure Beikenstein-Hawking area entropy, such as topological black hole \cite{topological,Nobuyoshi,Hamilton}, Kehagias-Sfetsos black hole \cite{adMyung,LML1}, Park black hole \cite{HLnewsolutionPark}, $z = 4$ black holes \cite{Rgcz4,Nobuyoshi} and so on.
For the (3 + 1) dimensional case of $z = 4$ HL gravity, the modifications are contained the logarithmic term and the terms of $1/A$ and $1/{A}^2$ in the entropy $S_{BH(3 + 1)}$ (\ref{shi11}). Furthermore, for the (4 + 1) dimensional case, there even is no any area term, but the terms of $A^{3/2}$ and $A^{1/2}$ in the entropy $S_{BH(4 + 1)} $ (\ref{equation17}). Hence, the thermodynamical entropy is no longer the pure spectrum of area. Through WP tunneling method, the tunneling entropies $S_{B(D + 1)}$, i.e. Eqs.(\ref{equation32}) and (\ref{equation48}), are the same as the thermodynamical entropies $S_{BH(D + 1)}$, i.e. Eqs.(\ref{shi11}) and (\ref{equation17}). The results suggest these non-area entropies could be treated as the self-gravitational effects of Hawking radiation, which is independent on radiating particle's mass. The first law of thermodynamics is still validity viewing the tunneling picture.

2. Comparing with the entropy of (3 + 1) dimensional black hole in the HL gravity corresponding $z = 3$ in the UV region constructed by the topological massive theory (TMG) \cite{topological,Nobuyoshi,Hamilton}, our corresponding results of (3 + 1) dimensional black hole entropy, i.e. $S_{B(3 + 1)}$ Eq.(\ref{equation32}), have some corrected term, i.e. $1/A$ and $1/{A}^2$ in the entropy of the $z = 4$ HL gravity constructed by the new massive gravity (NMG). There are sharp differences of results between the NMG theory and TMG theory. In (3 + 1) dimensions, as $\beta\rightarrow0$, the solution corresponding $z = 4$ in the UV region can reduce to the  $z = 3$ topological solution \cite{topological}. Meanwhile, the entropy at the horizon of the (3 + 1) dimensional black hole in the  $z = 4$ HL gravity also returns to the entropy in the $z = 3$ HL gravity, this implies NMG is an extent of the topological massive theory (TMG) gravity in some aspect. Else, it might also be noted that Hawking radiation of fermions are studied by Chen-Yang-Zu in Ref.\cite{Chendeyou} where the tunneling rate is determined by Hawking temperature, which is different from our work clearly. So far, we can conclude that the strange entropy of black hole appeared in $z = 4$ HL gravity could be explained well by the quantum tunneling with self-gravitational interaction.

\acknowledgments  Project is supported by the Natural Science Foundation of China (No.11005088), Foundation for University Key Teacher by He'nan Educational Committee, Foundation of He'nan Educational Committee (No.2011A140022). Jianbo Lv's work is supported by Natural Science Foundation of China (No.11205078). Yabo Wu's work is supported by Natural Science Foundation of China (No.11175077). Rumin Wang's work is supported by Natural Science Foundation of China (No.11105115).

\end{document}